\documentclass[10pt,a4paper]{article}

\usepackage{times}
\usepackage[centertags]{amsmath}
\usepackage{amssymb,amsfonts,theorem,stmaryrd}
\usepackage{slashbox}
\usepackage{pstricks}

\allowdisplaybreaks

\theorembodyfont{\slshape}
\newtheorem{thm}{Theorem}[section]
\newtheorem{prop}[thm]{Proposition}

\newtheorem{cor}[thm]{Corollary}
\theorembodyfont{\rmfamily} 

\newtheorem{rem}[thm]{Remark}

\newenvironment{pf}{\textit{Proof. }}{\hfill$\boxbox$}

\newcommand{\T}{\mathcal{T}}
\newcommand{\SC}{\mathcal{S}}
\newcommand{\V}{\mathcal{V}}

\newcommand{\Riem}{\operatorname{Riem}}

\newcommand{\sgn}{\operatorname{sign}}

\newcommand{\id}{\operatorname{id}}

\newcommand{\Tr}{\operatorname{Tr}}
\newcommand{\dvol}{\operatorname{dvol}}

\newcommand{\divergenz}{\operatorname{div}}

\newcommand{\bb}{\begin{eqnarray}}
\newcommand{\ee}{\end{eqnarray}}
\newcommand{\eee}{\nonumber\end{eqnarray}}

\begin{document}
\begin{center}
{\LARGE
The Holst Action by the Spectral Action Principle}\medskip

{\large Frank Pf\"aff\-le\footnote{ Email: pfaeffle@math.uni-potsdam.de} \& Christoph A.~Stephan\footnote{ Email: christophstephan@gmx.de}}\medskip

Institut f\"ur Mathematik\\
Universit\"at Potsdam\\
Am Neuen Palais 10\\
14469 Potsdam, Germany
\end{center}


\begin{abstract}
\noindent
We investigate the Holst action for closed Riemannian $4$-manifolds with orthogonal connections.
For connections whose torsion has zero Cartan type component 
we show that the Holst action can be recovered from the heat asymptotics for the natural Dirac operator acting on left-handed spinor fields.
\end{abstract}


\section{Introduction}
Connes' spectral action principle (\cite{Connes96}) states that any reasonable physical action should be deducible from the spectrum of some suitable Dirac operator.
One of the impressive achievements of the spectral action principle is the Chamseddine-Connes spectral action (\cite{ConnesChamseddine1}) which comprises the Einstein-Hilbert action of general relativity and the bosonic part of the action of the standard model of particle physics. It gives a conceptual explanation for the Higgs potential in order to have the electro-weak symmetry breaking, and  it allows to put constraints on the mass of the Higgs boson.
The present article is intended to show how the spectral action principle can be used to derive the Holst action.\medskip

\noindent
Loop Quantum Gravity (LQG) is a very promising and successful candidate for a theory of quantum gravity (see~\cite{Rovelli} and \cite{Thiemann} for an introduction).
Important ingredients of the quantisation procedure are the canonical variables of Ashtekar type (\cite{Ashtekar1}, \cite{Ashtekar2}).
In order to have such variables one considers the Holst action (\cite{Holst}) which is a modification of the Einstein-Cartan-Hilbert action with the same critical points.
Recently, a large class of modified actions with the same critical points as the Einstein-Cartan-Hilbert action has been proposed (\cite{DuboisViolette}).
This raises the question if the Holst action is distinguished in this large class of actions.
We will see how the spectral action principle gives a conceptual explanation for the Holst action for closed Riemannian $4$-manifolds, not only within the class of actions proposed in \cite{DuboisViolette}.
\medskip

\noindent  
The underlying geometric objects we will consider are orthogonal connections with general torsion in the sense of \'E.~Cartan (see \cite{Cartan23}, \cite{Cartan24}, \cite{Cartan25}).
(For an overview of the physical consequences of Einstein-Cartan theory
in the Lorentzian setting we refer to \cite{HHKN76} and \cite{S02}.)
The torsion of any orthogonal connection decomposes into a vectorial component, a totally anti-symmetric one and one of Cartan type.
In section~\ref{section1} we will recapitulate this decomposition and impose it into the Holst action in order to discuss the appearance of critical values of the Barbero-Immirzi parameters.
In section~\ref{section2} we will consider the classical Dirac operator $D$ associated to such an orthogonal connection.
The Cartan type component of the torsion does not affect $D$, and $D$ is not symmetric if the vectorial component of the torsion is non-zero.
We will derive a Lichnerowicz formula for $D^*D$ and deduce the heat trace asymptotics for the restriction of $D^*D$ to the left-handed spinor fields.
It turns out (in Cor.~\ref{spectralholst}) that the second term in these asymptotics gives exactly the Holst action if one considers connections with zero Cartan type torsion.
Furthermore the constraints from the spectral action principle allow us to fix the 
value of the Barbero-Immirzi parameter (\cite{Barbero}, \cite{Immirzi}) 
which is a free parameter in LQG.

\section{The Einstein-Cartan-Hilbert action and the Holst action}\label{section1}

For the convenience of the reader let us briefly recall the classical Cartan classification of orthogonal connections (see \cite[Chap.~VIII]{Cartan25}), we will adopt the notations of \cite[Chap.~3]{Tricerri}:
We consider an $n$-dimensional manifold $M$ equipped with some Euclidean metric $g$ and some orientation (in order to have a volume form and a Hodge $\ast$-operator).
Let $\nabla^{g}$ denote the Levi-Civita connection on the tangent bundle.
For any affine connection $\nabla$ on the tangent bundle there exists a $(2,1)$-tensor field $A$ such that
\begin{equation}\label{orthconnection}
\nabla_X Y= \nabla^g_X Y +A(X,Y)
\end{equation}
for all vector fields $X,Y$.
We will require all connections $\nabla$ to be {\it orthogonal}, i.~e.~compatible with the scalar product given by the Euclidean metric $g$.
Therefore, one has $g\left( A(X,Y), Z\right) = -g\left( Y,A(X, Z)\right)$ for any tangent vectors $X,Y,Z\in T_pM $.
The induced $(3,0)$-tensor is given by
\begin{equation*}
A_{XYZ} =g\left( A(X,Y),Z \right).
\end{equation*}
Hence, the space of all possible torsion tensors on $T_pM$ is 
\begin{equation*}
\Lambda^1\otimes\Lambda^2=\left\{ A\in {\bigotimes}^3T^*_pM\; \big| \; A_{XYZ}=-A_{XZY}\quad\forall X,Y,Z\in T_pM \right\}.
\end{equation*}
It carries a natural euclidean scalar product, which reads for any orthonormal basis $e_1,\ldots,e_n$ of $T_pM$ as
\begin{equation}\label{tensorscalprod}
\left\langle A,B \right\rangle= \sum_{i,j,k=1}^n A_{e_i e_j e_k} B_{e_i e_j e_k},
\end{equation}
and the orthogonal group $O(n)$ acts on it by $(\alpha A)_{XYZ}=A_{\alpha^{-1}(X)\alpha^{-1}(Y)\alpha^{-1}(Z)}$.
The corresponding norm is $\|A \|^2 = \langle A,A \rangle $
Then, one has the following decomposition of $\Lambda^1\otimes\Lambda^2$ into irreducible $O(n)$-subrepresentations:
\[
\Lambda^1\otimes\Lambda^2\,=\,\V(T_pM)\,\oplus\, \T(T_pM)\,\oplus \,\SC(T_pM).
\]
This decomposition is orthogonal with respect to $\langle\cdot,\cdot\rangle$, and it is given by
\begin{eqnarray*}
\V(T_pM)&=& \left\{ A\in \Lambda^1\otimes\Lambda^2 \;\big|\; \exists V \mbox{ s.t. } \forall X,Y,Z:\, A_{XYZ}= 
g(X,Y) \, g(V,Z)-g( X,Z)\,g( V,Y)\right\}, \\
\T(T_pM)&=& \left\{ A\in \Lambda^1\otimes\Lambda^2 \;\big|\; \forall X,Y,Z:\,A_{XYZ}=-A_{YXZ}\right\}, \\
\SC(T_pM)&=& \left\{ A\in \Lambda^1\otimes\Lambda^2 \;\big|\; \forall X,Y,Z:\,A_{XYZ}+ A_{YZX}+ A_{ZXY}=0\mbox{ and } \sum_{a=1}^n A(e_a,e_a,Z)=0\right\}.
\end{eqnarray*}
The connections whose torsion tensor is contained in $\V$ are called {\it vectorial}.
Those whose torsion tensor is in $\T$ are called {\it totally anti-symmetric},
and those with torsion tensor in $\SC$ are called {\it of Cartan type}.\medskip

\noindent
From this decomposition we get that for any orthogonal connection $\nabla$ as in (\ref{orthconnection}) 
there exist a  vector field $V$, a $3$-form $T$ and a $(3,0)$-tensor field $S$ with $S_p\in \SC(T_pM)$ for any $p\in M$ such that
\begin{equation}\label{uniquedecomp}
A(X,Y)=g( X,Y) V- g(V,Y ) X + T(X,Y,\cdot)^\sharp + S(X,Y,\cdot)^\sharp,
\end{equation}
these $V,T,S$ are unique.
As usual $^\sharp:T^*_pM\to T_pM$ denotes the canonical isomorphism induced by $g$.
The scalar curvature of this orthogonal connection is 
\begin{equation}\label{scalarcurv}
R= R^g +2(n-1)\,\divergenz^{g}(V) -(n-1)(n-2)\,|V|^2 -\|T\|^2+\tfrac{1}{2}\,\|S\|^2
\end{equation}
where $R^g$ is the scalar curvature of $\nabla^g$ and $\divergenz^{g}$ denotes the divergence taken with respect to $\nabla^g$ (see e.g.~\cite[Lemma 2.5]{Torsion}).\medskip

\noindent
Let $(\theta^a)_{a=1,\ldots,n}$ denote the dual frame of $(e_a)_{a=1,\ldots,n}$, i.e.~$\theta^a(\cdot)=g(e_a,\cdot)$.
Then the volume form is $\dvol=\theta^1\wedge\ldots\wedge \theta^n$.
For $k$-forms there is a natural scalar product $\langle,\rangle_k$ such that the elements $\theta^{i_1}\wedge\ldots\wedge \theta^{i_k}$ with $i_1<\ldots<i_k$ form an orthonormal basis of $\Lambda^k(T_p^*M)$ (compare \cite[Def.~0.1.4]{Bleecker}) 
Furthermore we have the Hodge operator $\ast:\Lambda^k(T_pM)\to \Lambda^{(n-k)}(T_pM)$, and for $\omega,\eta\in \Lambda^k(T_p^*M)$ one has
$\omega\wedge \ast\eta=\langle \omega, \eta\rangle_k\,\dvol$.\medskip

\noindent
For $n=4$ and $k=2$ we have $\ast\ast=\id$, 
this decomposes the space of $2$-forms into the selfdual and the anti-selfdual ones\footnote{Note that in Lorentzian signature $\ast\ast=-\id$ and hence $\ast$ has eigenvalues $\pm i$.}:
 $\Lambda^2= \Lambda^2_+\oplus\Lambda^2_-$, where $\Lambda^2_\pm$ is the $\pm 1$-eigenspace of $\ast$.
The (anti-)selfdual component of $\SC$ is denoted by
 $\SC^\pm =\SC\cap (\Lambda^1\otimes \Lambda^2_\pm)$.
Thereby, we obtain the decomposition 
\begin{equation}\label{decomposit}
\Lambda^1\otimes\Lambda^2 = \V \oplus \T\oplus \SC^+ \oplus \SC^-
\end{equation}
which is orthogonal w.r.t.~the scalar product given in (\ref{tensorscalprod}) and decomposes 
the component $S$ from (\ref{uniquedecomp}) into $S=S_+ +S_-$. \medskip

\noindent
In LQG (see \cite{Rovelli} or \cite{Thiemann}) one considers the case $n=4$, and most of the local computations are done in Cartan's moving frame formalism.
Let $(e_a)_{a=1,\ldots,4}$ be a local positively oriented orthonormal frame of $TM$ and $(\theta^a)_{a=1,\ldots,4}$ its dual frame.
To each orthogonal connection $\nabla$ as above one associates the connection $1$-forms $\omega_b^a$, the curvature $2$-forms $\Omega_b^a$ and the torsion $2$-forms $\Theta^a$, which are given by
\begin{eqnarray*}
\omega^a_b(X)&=& g(\nabla_X e_b,e_a), \\
\Omega^a_b(X,Y)&=& g(\Riem(X,Y) e_b,e_a)\; =\; g(\nabla_X\nabla_Ye_b-\nabla_Y\nabla_Xe_b-\nabla_{[X,Y]}e_b,e_a),\\
\Theta^a(X,Y)&=& g(A(X,Y)-A(Y,X),e_a)\; =\; g(\nabla_XY-\nabla_YX-[X,Y] ,e_a).
\end{eqnarray*}
For these forms one has the following structure equations:
\begin{eqnarray*}
\Omega^a_b&=& d\omega^a_b + \sum_{c} \omega^a_c\wedge \omega^c_b\\
\Theta^a &=& d\theta^a+\sum_{c} \omega_c^a\wedge \theta^c.
\end{eqnarray*}
With respect to the given frame one defines the {\it translational} Chern-Simons form by
\[
C_{TT}= \sum_a \Theta^a\wedge \theta^a.
\]
With the structure equations one obtains the Nieh-Yan equation (see \cite{NiehYan}):
\begin{equation}\label{NiehYan}
dC_{TT} = \sum_a \Theta^a \wedge \Theta^a + \sum_{a,b} \Omega^a_b\wedge\theta^b\wedge\theta^a.
\end{equation}
\begin{prop}
One has $C_{TT}=6T$ where $T$ is the totally anti-symmetric component of the torsion as in (\ref{uniquedecomp}).
\end{prop}
\pf{
We compute
\begin{eqnarray*}
\sum_a \Theta^a\wedge \theta^a (e_k,e_\ell,e_m)&=&  
\sum_a \left( 
\Theta^a(e_k ,e_\ell)\delta_{am}+  \Theta^a( e_\ell,e_m)\delta_{ak}+  \Theta^a(e_m ,e_k)\delta_{a \ell}
\right)\\
&=&  \Theta^m(e_k ,e_\ell)+  \Theta^k( e_\ell,e_m)+  \Theta^\ell(e_m ,e_k) \\
&=& \sum_\sigma \sgn(\sigma) A(e_{\sigma(k)},e_{\sigma(\ell)},e_{\sigma(m)})
\end{eqnarray*}
where the last summation is taken over all permutations of $\{k,\ell,m\}$, 
it is the anti-symmetrisation of $A\in\Lambda^1\otimes\Lambda^2$ and therefore equals $6T$.
{\hfill$\boxbox$}}\medskip

\noindent
This shows in particular that $C_{TT}$ is globally defined, i.e.~independent of the choice of the moving frame.
Sometimes it is stated that $\int_M dC_{TT}$ is a topological invariant, and it is called  Nieh-Yan invariant. 
\begin{cor}
Assume that the totally anti-symmetric component $T$ of the torsion has compact support which avoids the boundary of $M$, 
then
\[
\int_M dC_{TT} =0.
\]
\end{cor}
For the case when $M$ is closed the corollary was already shown in \cite{GuoWuZhang} by means of Chern-Weil theory.
It also holds in the Lorentzian setting and was already implicitly used e.g.~in \cite{DuboisViolette}.
If the support of $M$ meets the boundary Stoke's Theorem gives simple formulas for $\int_M dC_{TT}$.
Such terms have been considered e.g.~in \cite{Banerjee}.\medskip

\noindent
The Nieh-Yan equation is remarkable since the second summand equals the density of the {\it Holst term} (see \cite{Holst})
\begin{equation}
C_H=\sum_{a,b} \Omega^a_b\wedge\theta^b\wedge\theta^a.
\end{equation}
\begin{prop}\label{prop1}
For any orthogonal connection $\nabla$ one finds 
\[
C_H=6\,dT- 12\,\langle T,\ast V^\flat\rangle_3\,\dvol-\tfrac12\left(\|S_+\|^2-\|S_-\|^2\right)\dvol
\]
with respect to the decomposition (\ref{decomposit}), where $V^\flat(\cdot)=g(V,\cdot)$ is the dual form of the vector field $V$.
\end{prop}
\pf{
To simplify notation we abbreviate 
$\theta^{ab}=\theta^a\wedge\theta^b$, $\theta^{abc}=\theta^a\wedge\theta^b\wedge\theta^c$ and set
$A_{abc}=A(e_a,e_b,e_c)$, 
and likewise for all components of the decomposition (\ref{decomposit}), e.g.~$T_{abc}=T(e_a,e_b,e_c)$,
and we set $V_d=g(V,e_d)$.
We define the $2$-forms $A^a=\frac12\sum_{b,c} A_{abc}\theta^{bc}$ and likewise $V^a,T^a$ etc.
From the definition of $\Theta^a$ we get
\begin{eqnarray*}
\Theta^a(e_b,e_c)&=& T_{bca}+V_{bca}+S_{bca}-T_{cba}-V_{cba}-S_{cba}\\
                 &=& 2 T_{abc}-V_{abc}-S_{abc}.
\end{eqnarray*}
Therefore $\Theta^a=2T^a-V^a-S^a$, in the following we will consider the terms occuring 
\[
\sum_a\Theta^a\wedge\Theta^a = \sum_a 
\left(4\,T^a\wedge T^a 
-4\, T^a\wedge S^a 
-4\, T^a\wedge V^a 
+ V^a\wedge V^a
+ 2\,V^a\wedge S^a
+ S^a\wedge S^a\right).
\] 
For the first three terms we calculate
\begin{eqnarray}
4 \sum_a \,T^a\wedge A^a &=& \sum_{a,b,c,b',c'} T_{abc}A_{ab'c'}\;
 \theta^{bcb'c'}  \nonumber\\
&=&\sum_{a,b,c,d} T_{abc} \left(A_{aad}\theta^{bcad}  + A_{ada}\theta^{bcda} \right) \nonumber\\
&=& 2\,\sum_{a,b,c,d} T_{abc} A_{aad}\theta^{abcd}.\label{above}
\end{eqnarray}
For the second equality we observe that $\theta^{bcb'c'}\ne 0$ only if $b,c,b',c'$ are pairwise distinct and 
$T_{abc}\ne 0$ only if $a,b,c$ are pairwise distinct.
As $1\le a,b,c,b',c' \le 4$ only summands with $a=b'$ or $a=c'$ can contribute.\medskip

\noindent
For $A=T$ we get from  (\ref{above}) that $\sum_a \,T^a\wedge T^a=0$.
For $A=S$ we convince ourselves that $\sum_a \,T^a\wedge S^a=0$ by considering the sum in (\ref{above}) with fixed $d$, for example $d=4$:
\begin{eqnarray*}
\sum_{a,b,c} T_{abc} S_{aa4}\theta^{abc4}&=& 
S_{114}(T_{123}\theta^{1234}+T_{132}\theta^{1324})+
S_{224}(T_{213}\theta^{2134}+T_{231}\theta^{2314})+
S_{334}(T_{312}\theta^{3124}+T_{321}\theta^{3214})
\\
&=& 2\,T_{123}\left( S_{114} + S_{224} + S_{334}\right)\theta^{1234}\quad = \quad 0
\end{eqnarray*}
since $S_{444}=0$ and the trace of $S$ over the first two entries vanishes.
For $A=V$ we have $A_{aad}=V_d-\delta_{ad}V_a$, which we insert into (\ref{above})
\begin{eqnarray*}
-4 \sum_a \,T^a\wedge V^a &=& -2\,\sum_{a,b,c,d} T_{abc} V_d\theta^{abcd}\\
&=& -2 \,\big( \sum_{a,b,c} T_{abc}\theta^{abc}\big)\wedge \big( \sum_d V_d\theta^d\big)\\
&=& -12 \,T\wedge V^\flat \quad =\quad 12\,\langle T,\ast V^\flat\rangle_3\,\dvol.
\end{eqnarray*}
Similarly, we obtain
\begin{equation}\label{above2}
\sum_a \,V^a\wedge A^a\;=\;-\tfrac12 \,\sum_{a,b,c,d} V_aA_{bcd}\theta^{abcd},
\end{equation}
which is zero for $A=V$. 
In the case of $A=S$ in (\ref{above2}) we notice
\[
-\sum_{a,b,c,d} V_aS_{bcd}\theta^{abcd}= \sum_{a,b,c,d} V_a\big( S_{cdb}+S_{dbc}\big)\theta^{abcd} 
= 2\, \sum_{a,b,c,d} V_aS_{bcd}\theta^{abcd} = 0.
\]
Finally, with $\ast S^a=S^a_+-S^a_-$ and $\sum_a\langle S_{(\pm)}^a, S_{(\pm)}^a\rangle_2 =\frac12 \| S_{(\pm)}\|^2$ we get
\[
\sum_a S^a\wedge S^a= \sum_a \langle S^a,\ast S^a\rangle_2\,\dvol 
= \tfrac12\left( \|S^+\|^2- \|S^-\|^2\right)\,\dvol \,.
\]
We conclude that $\sum_a \Theta^a\wedge\Theta^a = 12\,\langle T,\ast V^\flat\rangle_3\,\dvol+\tfrac12\left(\|S^+\|^2- \|S^-\|^2\right)\,\dvol$.
With the Nieh-Yan equation (\ref{NiehYan}) and Prop.~\ref{prop1} the claim follows.
{\hfill$\boxbox$}}\medskip

\noindent
This shows that $C_H$ depends only on the torsion of the connection but not on the Riemannian curvature of the underlying manifold.
Observations of that kind have been made before in \cite{HM86}, \cite{Mercuri} and \cite{Banerjee}.\medskip

\noindent
The {\it Holst action}\footnote{Before \cite{Holst} this action already appeared in \cite{HMS80}, a sketch of its history can be found in \cite[Section III.D]{BHN11}.} used in LQG is given by
\begin{equation*}
\mathcal{I}_H =\frac{1}{16\pi G}\int_M \rho_\gamma^\nabla\,\dvol=\frac{1}{16\pi G}\int_M \left(R\,\dvol-\frac1\gamma \,C_H\right),
\end{equation*}
where $G$ is Newton's constant and $\gamma$ is the Barbero-Immirzi parameter (\cite{Barbero}, \cite{Immirzi}).
The density of the Holst action reads as
\begin{eqnarray}
\rho_\gamma^\nabla\,\dvol&=& \left(R^g +6\,\divergenz^{g}(V) -6\,|V|^2 -\|T\|^2 + \tfrac{12}{\gamma}\langle T,\ast V^\flat\rangle_3\right)\,\dvol-\tfrac{6}{\gamma}\, dT \nonumber\\ 
&&\; +\left(\tfrac{1}{2}(1+\tfrac1\gamma )\,\|S^+\|^2+\tfrac{1}{2}(1-\tfrac1\gamma )\,\|S^-\|^2\right)\,\dvol\label{holstaction}
\end{eqnarray}
Assuming that both $T$ and $V$ have compact support and avoid the boundary of $M$ we obtain
\[
\mathcal{I}_H =\frac{1}{16\pi G}\int_M \left( R^g  -6\,|V|^2 -\|T\|^2+\tfrac{12}{\gamma}\langle T,\ast V^\flat\rangle_3  +\tfrac{1}{2}(1+\tfrac1\gamma )\,\|S^+\|^2+\tfrac{1}{2}(1-\tfrac1\gamma )\,\|S^-\|^2
\right)\dvol.
\]
If $\gamma=\pm1$ one can vary $S^+$ or $S^-$ without changing the value of $\mathcal{I}_H$, thus obtaining more critical points than for the Einstein-Hilbert functional.\footnote{In Lorentzian signature the critical values of the Barbero-Immirzi parameters are $\pm i$.}
These critical values of the Barbero-Immirzi are well known in LQG, we think our representation of the Holst action offers a clear geometric understanding of this fact.

\section{Dirac operators and the spectral action principle}\label{section2}

The spectral action principle (\cite{Connes96}) of noncommutative geometry (\cite{Connes94}) states that the whole information of physical reality is encoded in  some suitable Dirac operator, and one should be able to extract any measurable quantity from its spectrum.
In the following we want to discuss the relation between the classical Dirac operator and the Holst action.\medskip

\noindent
We consider an $n$-dimensional Riemannian manifold
and we assume that $M$ carries a spin structure so that spinor fields are defined.
Any orthogonal connection $\nabla$ as in (\ref{orthconnection}) induces a unique connection acting on spinor fields (see \cite[Chap.~II.4]{LawsonMichelsohn} or \cite[Section~4]{Torsion}) which we will also denote by $\nabla$.
The Dirac operator associated to $\nabla$ is defined as
\begin{eqnarray}
D\psi = \sum_{a=1}^n e_a\cdot \nabla_{e_a}\psi 
&=& D^g\psi +\tfrac14\sum_{a,b,c=1}^n  A_{abc} \, e_a\cdot e_b \cdot e_c \cdot \psi \nonumber\\
&=& D^g\psi +\tfrac32 T\cdot \psi -\tfrac{n-1}{2}V\cdot \psi\label{def_torsiondirac}
\end{eqnarray}
where $D^g$ is the Dirac operator associated to the Levi-Civita connection and ``$\cdot$'' is the Clifford multiplication.\footnote{For the Clifford relations we use the convention $X\cdot Y+Y\cdot X=-2\,g(X,Y)$ for any tangent vectors $X,Y$, and any $k$-form $\theta^{i_1}\wedge\ldots\wedge\theta^{i_k}$ acts on some spinor $\psi$ by $\theta^{i_1}\wedge\ldots\wedge\theta^{i_k}\cdot\psi = e_{i_1}\cdot\ldots \cdot e_{i_k}\cdot \psi$.}
Using the fact that the Clifford multiplication by the vector field $V$ is skew-adjoint w.r.t.~the hermitian product on the spinor bundle
one observes that  that $D$ is symmetric with respect to the natural $L^2$-scalar product on spinors if and only if the vectorial component of the torsion vanishes, $V\equiv 0$ (see \cite{FriedrichSulanke} and \cite{Torsion}, and \cite{GS87} for the Lorentzian setting).
We would like to stress that the Dirac operator $D$ stays pointwise the same if one changes the Cartan type component $S$ of the torsion (see e.g.~\cite[Lemma~4.7]{Torsion}).\footnote{In the Lorentzian case it is known that torsion of Cartan type does not 
contribute to the Dirac action under the integral \cite[Chap.~2.3]{S02}.}
Therefore the Dirac operator $D$ does not contain any information on the Cartan-type component $S$ of the torsion.
We summerise:
\begin{cor}
In general, neither $C_H$ nor $\int_M C_H$ nor $\mathcal{I}_H$  can be recovered from the spectrum of the $D$.{\hfill$\boxbox$}
\end{cor}
\begin{rem}
For any compact spin manifold the Atiyah-Singer Index Theorem relates the index of the left-handed Dirac operator (mapping left-handed to right-handed spinor fields) to a topological invariant of the manifold (the $\hat{A}$-genus).
The index of an elliptic operator depends only on its principal symbol (see e.g.~\cite[Cor.III.7.9]{LawsonMichelsohn}).
Therefore, the index of the left-handed part of Dirac operator defined in (\ref{def_torsiondirac}) does not depend on the torsion. 
The index density in the case of ``$H$-torsions'' (i.e.~totally anti-symmetric torsion with $dT=0$) has been calculated in \cite{Kimura}.
\end{rem}
Nevertheless, we will recover the Holst action from the  heat trace asymptotics for $D^*D$ if we restrict to the case $S\equiv 0$, which is the natural case when dealing with spinors.
First, we derive the following Lichnerowicz formula:
\begin{thm}
For the Dirac operator $D$ associated to the orthogonal connection $\nabla$ as given in (\ref{def_torsiondirac}) we have
\begin{eqnarray}\label{Lichnerowicz}
D^*D\psi &=& \Delta\psi + \tfrac14 R^g\,\psi + \tfrac32 dT\cdot \psi -\tfrac34\|T\|^2\,\psi\nonumber \\
&& \; + \tfrac{n-1}{2}\divergenz^g(V)\,\psi +\left(\tfrac{n-1}{2}\right)^2(2-n)\,|V|^2\,\psi \nonumber \\
&& \; +3(n-1)\Big(T\cdot V\cdot\psi +(V\lrcorner T)\cdot\psi \Big)
\end{eqnarray}
for any spinor field $\psi$, where $\Delta$ is the Laplacian associated to the connection 
\[
\widetilde{\nabla}_X\psi = \nabla^g_X\psi +\tfrac32 (X\lrcorner T)\cdot \psi - \tfrac{n-1}{2}V\cdot X\cdot \psi - \tfrac{n-1}{2}\, g(V, X)\, \psi.
\]
\end{thm}
\pf{ As Clifford multiplication by any $3$-form is self-adjoint we have
\[
D^*\psi=D^g\psi +\tfrac32 T\cdot \psi +\tfrac{n-1}{2}V\cdot \psi.
\]
We calculate
\begin{eqnarray}
D^*D\psi&=& \left(D^g+\tfrac32 T\cdot \right)^2\psi - \left(\tfrac{n-1}{2}\right)^2 \,V\cdot V\cdot\psi \nonumber\\
&&\; +\,\tfrac{3\,(n-1)}{4}\left(V\cdot T -T\cdot V \right)\cdot\psi +\tfrac{n-1}{2}\left(V\cdot D^g - D^g V\cdot \right)\psi\nonumber\\
&=& \left(D^g+\tfrac32 T\cdot \right)^2\psi + \left(\tfrac{n-1}{2}\right)^2 \,|V|^2\,\psi \nonumber\\
&&\; -\,\tfrac{3\,(n-1)}{2}\big( T\cdot V+(V\lrcorner T) \big)\cdot\psi
+\tfrac{n-1}{2}\left(2V\cdot D^g\psi +2\nabla^g_V\psi +\divergenz^g(V)\,\psi-d(V^\flat)\cdot\psi\right)\qquad \label{above3}
\end{eqnarray}
where we have used the relation $V\cdot T +T\cdot V=-2(V\lrcorner T)$ for the vector $V$ and the $3$-form $T$ and the identity 
$D^g V+VD^g = -2\nabla^g_V-\divergenz^g(V)+d(V^\flat)$.
In order to calculate the Laplacian associated to the connection $\widetilde{\nabla}$ we fix some $p\in M$ and choose the frame $(e_a)$ to be synchronous about $p$, i.e.~$\nabla^ge_a|_p=0$ for any $a=1,\ldots,n$.
\begin{eqnarray*}
\Delta \psi&=& -\sum_a \widetilde{\nabla}_{e_a}\widetilde{\nabla}_{e_a}\psi\\
&=& - \sum_a\left(\nabla^g_{e_a}+\tfrac32\,(e_a\lrcorner T) \right)\left(\nabla^g_{e_a}+\tfrac32\,(e_a\lrcorner T) \right)\psi \\
&&\; + \sum_a \tfrac{n-1}{2} \left(\nabla^g_{e_a}+\tfrac32\,(e_a\lrcorner T) \right)\left(V\cdot e_a+g(V,e_a) \right)\psi \\
&&\; + \sum_a \tfrac{n-1}{2} \left(V\cdot e_a+g(V,e_a) \right)\left(\nabla^g_{e_a}+\tfrac32\,(e_a\lrcorner T) \right)\psi \\
&&\; - \sum_a \left(\tfrac{n-1}{2}\right)^2 \left(V\cdot e_a+g(V,e_a) \right) \left(V\cdot e_a+g(V,e_a)\right)\psi \\
&=& \left(D^g+\tfrac32 T\cdot \right)^2\psi- \tfrac14 R^g\,\psi - \tfrac32 dT\cdot \psi +\tfrac34\|T\|^2\,\psi\nonumber \\
&& \; + \tfrac{n-1}{2}\left( V\cdot D^g\psi +\nabla^g_V\psi +\tfrac32\,(V\lrcorner T)\cdot\psi -d(V^\flat)\cdot\psi +\tfrac32\sum_a (e_a\lrcorner T)\cdot V\cdot e_a\cdot\psi\right)\\
&& \; + \tfrac{n-1}{2}\left( V\cdot D^g\psi +\nabla^g_V\psi +\tfrac32\,(V\lrcorner T)\cdot\psi  +\tfrac32\sum_a V\cdot e_a\cdot(e_a\lrcorner T)\cdot \psi\right)\\
&& \,-\left(\tfrac{n-1}{2}\right)^2(1-n)|V|^2\,\psi,
\end{eqnarray*}
here we have used $\left(D^g+\tfrac32 T \right)^2 = - \sum_a\left(\nabla^g_{e_a}+\tfrac32\,(e_a\lrcorner T) \right)\left(\nabla^g_{e_a}+\tfrac32\,(e_a\lrcorner T) \right)+\tfrac{1}{4}R^g +\tfrac32 dT-\tfrac{3}{4}\|T \|^2$ which is Thm.~6.2
of \cite{AgricolaFriedrich} adapted to our notation.
Next we can deduce from $e_a\lrcorner T=\tfrac12\sum_{b,c}T_{abc}\theta^{bc}$ that $3T= \sum_a (e_a\lrcorner T)\cdot e_a=\sum_a e_a\cdot(e_a\lrcorner T)$ and we further simplify:
\begin{eqnarray*}
\Delta \psi&=& \left(D^g+\tfrac32 T\cdot \right)^2\psi- \tfrac14 R^g\,\psi - \tfrac32 dT\cdot \psi +\tfrac34\|T\|^2\,\psi\\
&&\; +\tfrac{n-1}{2}\left( 2\,V\cdot D^g\psi +2\,\nabla^g_V\psi  -d(V^\flat)\cdot\psi-9(V\lrcorner T)\cdot\psi- 9T\cdot V\cdot \psi \right) +\tfrac{(n-1)^3}{4}|V|^2 \psi.
\end{eqnarray*}
Together with (\ref{above3}) this yields the claim.
{\hfill$\boxbox$}}\medskip

\noindent
Now, let $k_t(x,y)$ denote the (smooth) kernel of the heat operator $\exp(-t D^*D)$.
Then one has the well-known asymptotic expansion 
\[
k_t(x,x) \sim \tfrac{1}{(4\pi t)^2}\left( \alpha_0(x) + t\,\alpha_2(x) +t^2\,\alpha_4(x)+\ldots \right)\quad\mbox{ for }t\to 0
\]
with $\alpha_0(x)=\id$.\medskip

\noindent
We have $\gamma_5=e_1\cdot\ldots\cdot e_4=\dvol$ and the projection on left-handed spinors is given by $P_{L}=\frac12(\id-\gamma_5)$.
Now let us consider the restriction $P_LD^*DP_L$ to the left-handed spinors.
We note that its kernel is given by $p_t(x,y)=P_L \circ k_t(x,y)\circ P_L$.
Taking the trace over the ($4$-dimensional) spinor spaces we obtain the following asymptotics
\begin{equation}\label{traceasympt}
\Tr \left(p_t(x,x) \right) \,\sim \,\tfrac{1}{(4\pi t)^2}\left(\beta_0(x)+ t\,\beta_2(x)+t^2\,\beta_4(x)+\ldots \right)
\quad\mbox{ for }t\to 0.
\end{equation}
As $\alpha_0(x)=\id$ we have $\beta_0(x)=2$ for any $x\in M$,
and the function $\beta_2(x)$ is related to the the density of the Holst action from (\ref{holstaction}) as follows.
Restricting to orthogonal connections with $S\equiv 0$ is natural since fermions are not able to perceive any torsion of Cartan type.
\begin{thm}\label{satz}
Let $M$ be a compact Riemannian $4$-manifold with spin structure.
For a $3$-form $T$ and a vector field $V$ consider the orthogonal connection $\nabla_XY=\nabla^g_XY+T(X,Y,\cdot)^\sharp + g(X,Y)V -g(V,Y)X$.
Let $D$ denote the Dirac operator induced by $\nabla$, and consider the restriction $P_LD^*DP_L$ to the left-handed spinors.
Then, for the term $\beta_2$  from the expansion (\ref{traceasympt}) we have
\begin{equation}\label{hauptergebnis}
\beta_2\,\dvol = -\tfrac16\,\rho_\gamma^{\overline{\nabla}}\dvol 
\end{equation}
for the orthogonal connection $\overline{\nabla}$ given by  $\overline{\nabla}_XY=\nabla^g_XY+3T(X,Y,\cdot)^\sharp + 3g(X,Y)V - 3g(V,Y)X$ and for the value $\gamma=1$ of the Barbero-Immirzi parameter. 
\end{thm}
\pf{
We use the Lichnerowicz formula (\ref{Lichnerowicz})  and the explicit formula for $\alpha_2(x)$ from e.~g.~\cite[Prop.~7.19]{Roe} to get
\[
\alpha_2=\tfrac16 R^g-\tfrac14 R^g - \tfrac32 dT +\tfrac34\|T\|^2 - \tfrac{3}{2}\divergenz^g(V) +\tfrac92 |V|^2-9 \left(T\cdot V+ V\lrcorner T \right).
\]
By construction one has $\beta_2= \tfrac12\Tr \left((1-\gamma_5)\alpha_2 \right)$.
We observe that the traces of $dT$, $T\cdot V$ and $V\lrcorner T$, taken over the $4$-dimensional spinor space, all vanish since they act as $2$-forms or $4$-forms.
So we get
\[
\Tr(\alpha_2) = -\tfrac13 R^g + 3\|T\|^2 -6\divergenz^g(V)+18|V|^2.
\]
For $a<b<c$ and $a'<b'<c'$ we get $\Tr(e_ae_be_ce_{a'}e_{b'}e_{c'})=4$ if $a=a'$, $b=b'$ and $c=c'$, and otherwise $\Tr(e_ae_be_ce_{a'}e_{b'}e_{c'})=0$.
Hence, for $3$-forms $T,T'$ we have
$\Tr(TT')=4\langle T,T' \rangle_3$.
We remark that $V\gamma_5=-\ast V^\flat$ and so $\Tr(T\cdot V\gamma_5)=-4 \langle T,\ast V^\flat \rangle_3$.
Furthermore, we have $\Tr(dT\,\gamma_5)\dvol =4\,dT$ and $\Tr(V\lrcorner T\,\gamma_5)\dvol =0$. 
This leads to
\[
\Tr(\gamma_5\alpha_2)\dvol = - 6 \,dT +36 \langle T,\ast V^\flat \rangle_3 \dvol.
\]
We obtain
\begin{eqnarray*}
\beta_2\,\dvol &=& \tfrac12\left( \Tr \left(\alpha_2 \right) - \Tr \left(\gamma_5\alpha_2 \right)\right)\,\dvol\\
&=& -\tfrac16 \left( 
\left(R^g - 9\|T\|^2 +18\divergenz^g(V)-54\,|V|^2 +108 \langle T,\ast V^\flat \rangle_3\right)\dvol - 18\,dT\right).
\end{eqnarray*}
Finally we compare this with the density of the Holst action $\rho_\gamma^{\overline{\nabla}}\dvol$ for the orthogonal connection $\overline{\nabla}$ and  the Barbero-Immirzi parameter $\gamma=1$, which is given by
\begin{equation*}
\rho_\gamma^{\overline{\nabla}}\dvol\; =\;\left(R^g - 9\|T\|^2 +18\divergenz^g(V)-54\,|V|^2 +108 \langle T,\ast V^\flat \rangle_3\right)\dvol - 18\,dT,
\end{equation*}
and establish (\ref{hauptergebnis}).
{\hfill$\boxbox$}}
\begin{cor}\label{spectralholst}
Let $M$ be a $4$-dimensional compact manifold and $\nabla$ be an orthogonal connection without Cartan type component as in Theorem~\ref{satz}.
Then we get for the second coefficient of the heat trace asymptotics for $P_LD^*DP_L$
\[
\int_M\beta_2\,\dvol\;=\; -\tfrac{1}{6} \int_M \rho_\gamma^{\overline{\nabla}}\dvol\; =\; -\tfrac{8\pi G}{3}\, \overline{\mathcal{I}}_H
\]
where $\overline{\mathcal{I}}_H$ denotes the Holst action for the connection $\overline{\nabla}$ with Barbero-Immirzi parameter $\gamma=1$.
{\hfill$\boxbox$}
\end{cor}
In other words Corollary~\ref{spectralholst} states that the spectral action principle naturally predicts the Holst action in the case of orthogonal connection without the Cartan type torsion (which is invisible to fermions).
The Barbero-Immirzi parameter then takes the critical value $\gamma=1$.
\medskip

\noindent
In~\cite{DuboisViolette} it has been proposed to modify the Holst action by adding terms depending on the norm of torsion $(\Theta,\Theta)=\sum_a \langle\Theta^a,\Theta^a \rangle_2$ and it is shown that such actions in general still have the same critial points as the Einstein-Cartan-Hilbert functional.
Apart from considerations of quantisation (the need of canonical variables of Ashtekar type) Corollary~\ref{spectralholst} shows that the Holst action is special within this proposed larger class of actions.
\medskip

\noindent
There has been a controversy whether the term $dC_{TT}=6dT$ could be obtained via anomaly calculations and what its significance in quantum field theory and its relevance for the Barbero-Immirzi parameter might be (\cite{ChandiaZanelli1}, \cite{ObukhovMielke}, \cite{KraimerMielke}, \cite{ChandiaZanelli12}).
In \cite{BrodaSzanecki} the induced gravity approach delivers the value $\pm\frac{1}{9}$ for the prefactor of the term $dC_{TT}$ if one addionally takes the specific particle content of the Standard Model into account.
Within the approach of Connes' spectral action principle a comparison of parameters would obtain the value $1$ for the  prefactor of the term $dC_{TT}$ (or the value $-1$ if we had projected on the right-handed spinors).
This value would be independent of any specific particle model.
However, one should be aware that in these actions the Cartan type component of the torsion does not appear, unlike in the action $\mathcal{I}_H$ which is considered to be the relevant one in LQG.
\medskip

\noindent
{\bf Acknoledgement:} The authors appreciate funding by the Deutsche Forschungsgemeinschaft, in particular by the SFB {\it Raum-Zeit-Materie}.
We would like to thank Christian B\"ar and Thomas Sch\"ucker for their support and helpful discussions.
Furthermore, we are thankful to Friedrich Hehl for drawing our attention to literature that was previously unknown to us.

%

\end{document}